\documentclass[a4paper,11pt]{article}
\usepackage{pos}
\usepackage{bbold}
\usepackage{graphicx}
\usepackage{physics}
\usepackage{subcaption}
\usepackage{cleveref}
\usepackage{amsmath}

\newcommand{\beq}{\begin{equation}}
\newcommand{\eeq}{\end{equation}}
\newcommand{\beqs}{\begin{eqnarray}}
\newcommand{\eeqs}{\end{eqnarray}}

\newcommand\Tab[1]{Table~\ref{tab:#1}}
\newcommand\Fig[1]{Fig.~\ref{fig:#1}}

\title{Spectroscopy of $Sp(4)$ lattice gauge theory with $n_f=3$ antisymmetric fermions}

\author*[a,b]{Jong-Wan Lee}
\author[c]{Ed Bennett}
\author[a]{Deog Ki Hong}
\author[d]{Ho Hsiao}
\author[d,e]{C.-J. David Lin}
\author[c,f]{Biagio Lucini}
\author[g]{Maurizio Piai}
\author[h]{Davide Vadacchino}

\affiliation[a]{Department of Physics, Pusan National University, Busan 46241, Korea}
\affiliation[b]{Institute for Extreme Physics, Pusan National University, Busan 46241, Korea}
\affiliation[c]{Swansea Academy of Advanced Computing, Swansea University (Bay Campus), Fabian Way, SA1 8EN Swansea, Wales, United Kingdom}
\affiliation[d]{Institute of Physics, National Yang Ming Chiao Tung University, 1001 Ta-Hsueh Road, Hsinchu 30010, Taiwan}
\affiliation[e]{Center for High Energy Physics, Chung-Yuan Christian University, Chung-Li 32023, Taiwan}
\affiliation[f]{Department of Mathematics, Faculty of Science and Engineering, Swansea University (Bay Campus), Fabian Way, SA1 8EN Swansea, Wales, United Kingdom}
\affiliation[g]{Department of Physics, Faculty of Science and Engineering, 
Swansea University (Park Campus), Singleton Park, SA2 8PP Swansea, Wales, United Kingdom}
\affiliation[h]{Centre for Mathematical Science, University of Plymouth, Plymouth, PL4 8AA, United Kingdom}

\emailAdd{jwlee823@pusan.ac.kr}
\emailAdd{e.j.bennett@swansea.ac.uk}
\emailAdd{dkhong@pusan.ac.kr}
\emailAdd{thepaulxiao@gmail.com}
\emailAdd{dlin@nycu.edu.tw}
\emailAdd{b.lucini@swansea.ac.uk}
\emailAdd{m.piai@swansea.ac.uk}
\emailAdd{davide.vadacchino@plymouth.ac.uk}

\abstract{We perform numerical calculations of masses and decay constants of the lightest (flavoured) pseudoscalar, vector and axial vector mesons in the $Sp(4)$ lattice gauge theory 
with three Dirac fermions in the antisymmetric representation. 
The corresponding continuum theory plays an important role in certain ultra-violet complete realisations of composite Higgs, partial top compositeness, and composite dark matter models. 
In addition, we measure the masses of other flavoured mesons in spin-$0$ and $1$ channels, as well as the first excited state of the vector mesons. 
Using the gradient flow to set the scale, we carry out the continuum extrapolation and show preliminary results 
for the meson spectrum of the theory. 
}%

\FullConference{%
The 39th International Symposium on Lattice Field Theory,\\
8th-13th August, 2022,\\
Rheinische Friedrich-Wilhelms-Universität Bonn, Bonn, Germany
}

\begin{document}
\maketitle

\section{Introduction}
In the context of physics beyond the standard model (BSM) based on novel strong dynamics, 
gauge theories with fermions in the two-index representation have great phenomenological potential, 
as they yield large coset spaces associated with the explicit and/or spontaneous breaking of the global flavour symmetry. 
Even with a small number of flavours, furthermore, it is expected that the theory might lie in or near the sill of the conformal window. 
Along the lines of a long-term research programme on nonperturbative lattice studies of $Sp(2N)$ gauge theories \cite{Bennett:2017kga}, 
we focus on the $Sp(4)$ lattice gauge theory coupled to three Dirac fermions in the antisymmetric (AS) representation. 

The global (flavour) symmetry breaking of the theory is $SU(6)\rightarrow SO(6)$ as the representation is real, 
and yields $20$ (massless) pseudo Nambu-Goldstone Bosons (pNGBs). 
A natural subgroup of $SU(6)$ is $SU(3)_L\times SU(3)_R$ and, after the symmetry breaking, 
the diagonal $SU(3)_D$ can be embedded in the unbroken subgroup of $SO(6)$. 
This feature is crucial to realise a mechanism of (top) partial compositeness by assigning $SU(3)$ colour charge to spin-$\frac{1}{2}$ baryon states in the new strong sector, 
such that these states can linearly couple to quarks in the standard model \cite{Barnard:2013zea,Ferretti:2013kya}. 
Although the complete model requires two additional fundamental Dirac fermions in the sea, 
its dynamics is well approximated by the $Sp(4)$ theory with only $n_f=3$ AS fermions. 
The theory can also provide an ultraviolet (UV) completion for a certain alternative scenario 
for composite Higgs and dark matter \cite{Cacciapaglia:2019ixa,Cai:2020njb}. 

In our contributions to previous lattice proceedings, we had reported some preliminary results about the bulk first-order phase transition \cite{Lee:2018ztv} 
and finite volume effects \cite{Lucini:2021xke} in the lattice theory formulated with the standard plaquette gauge and the Wilson-Dirac fermion actions.  
Since then, we have generated dynamical ensembles at various lattice couplings and (degenerate) fermion masses in moderate lattice volumes, 
which have been chosen such that finite volume effects are statistically negligible. 
By using these ensembles we measure the two-point correlation functions of spin-$0$ and $1$ flavoured meson operators 
and extract the masses and decay constants from the asymptotic behavior at late Euclidean time. 
After setting a common scale through the gradient flow method, 
we carry out the continuum extrapolation for the first time in this theory. 
In the case of vector and tensor mesons, which interpolate the same physical state with the quantum number of $J^P=1^-$, 
we extract the masses of both ground and first excited states. 

\section{Lattice Model}

Our lattice theory is built upon a discretised $4$-dimensional Euclidean space-time with the same gauge group and the same flavour structure of the continuum theory. 
For the gauge part we adopt the standard Wilson plaquette action and for the fermion part the Wilson-Dirac fermionic action, which are defined as
\beq
S\equiv \beta \sum_x\sum_{\mu<\nu} \left(1-\frac{1}{4}{\textrm Re}\,{\textrm Tr}\, U_\mu(x) U_\nu(x+\hat{\mu}) U_\mu^\dagger (x+\hat{\nu}) U_\nu^\dagger (x)\right) 
+ a^4 \sum_{j=1}^{n_f=3}\sum_{x,y} \overline{\Psi}^j(x) D_m (x,y) \Psi^j(y),
\eeq
where $\beta=8/g^2$ and $a$ are the lattice coupling and spacing, respectively. 
The massive Wilson-Dirac operator $D_m$ given by
\beq
D_m(x,y) = (4/a+m_0)\delta_{x,y} - \frac{1}{2a} \sum_\mu \left\{
(1-\gamma_\mu) U^{(as)}_\mu(x) \delta_{x,y-\hat{\mu}} + (1+\gamma_\mu) U^{(as),\,\dagger}_{\mu}(x) \delta_{x,y+\hat{\mu}}
\right\},
\eeq
where $m_0$ is the (degenerate) bare fermion mass. The fundamental link variable $U_\mu \in Sp(4)$ satisfies the condition, $U^*=\Omega U \Omega^\dagger$, 
with the symplectic matrix $\Omega = \begin{pmatrix}0 & 1 \\ -1 & 0\end{pmatrix}$. 
The antisymmetric link variable $U_\mu^{(as)}$ has been constructed following the prescription in Ref.~\cite{DelDebbio:2008zf} 
and subject to the $\Omega$-tracelessness condition \cite{Bennett:2019cxd}.
With this Wilson formulation the global symmetry is explicitly broken by both the mass and the Wilson terms 
at finite lattice spacing following the same pattern as the continuum theory. 
For the fermion fields we use periodic and antiperiodic boundary conditions in the spacial ($L$) and temporal ($T$) extents, respectively, 
while we use periodic boundary conditions in all the lattice extents for the gauge fields. 

We perform numerical simulations using the (rational) hybrid Monte Carlo ((R)HMC) algorithms implemented in the HiRep code \cite{DelDebbio:2008zf} 
with appropriate modifications \cite{Bennett:2017kga,Bennett:2019cxd,Bennett:2019jzz,Bennett:2022yfa}, 
where for the three Dirac flavours we use one pseudofermion for each of HMC and RHMC evolutions. 
Preliminary studies showed that the lattice theory undergoes a first-order bulk phase transition and the weak coupling regime is restricted to $\beta\gtrsim 6.5$ \cite{Lee:2018ztv}. 
Furthermore, it has been shown that finite volume effects are statistically negligible for $m_{\rm PS} \, L \gtrsim 8$ \cite{Lucini:2021xke}. 
Accordingly, for the ensemble generation we choose the lattice parameters satisfying these two conditions, 
with the lattice sizes from $48\times16^3$ to $54\times36^3$, such that lattice artefacts can systematically be studied. 
In particular, we consider four different beta values $\beta=6.65,\,6.7,\,6.75,\,6.8$, and several bare masses at each $\beta$. 
A common scale is set by the gradient flow method with the definition $w_0$ proposed in Ref.~\cite{Borsanyi:2012zs}, 
where we choose a specific reference value of $W_0=0.28$ suggested by the large-$N$ scaling behavior of the gradient flow scale \cite{Bennett:2022ftz}. 
We use hatted notation for any dimensionful quantities in units of $w_0$, e.g. $\hat{m}=m w_0$. 


\section{Observables}

\begin{table}[]
    \footnotesize
    \centering
    \caption{List of operators $\mathcal{O}_M$ interpolating meson states in the spin-$0$ and $1$ channels. 
    The operators are restricted to flavour non-singlets, $i\neq j$, and identified by the spin ($J$), parity ($P$), and flavour quantum numbers of the physical states. 
}
    \label{tab:meson_ops}
    \begin{tabular}{|c|c|c|c|c|}
    \hline
    ~~Label $M$ ~~& $\mathcal{O}_M$ & ~~~$J^P$~~~ & ~~$SO(6)$~~ & Meson in QCD\\
    \hline\hline
    PS & $\overline{\Psi^i}\gamma^5\Psi^j$ &  $0^-$ & 20 & $\pi$\\
    \hline
    S & $\overline{\Psi^i}\Psi^j$ & $0^+$ & 20 & $a_0$\\
    \hline
    V & $\overline{\Psi^i}\gamma^{\mu}\Psi^j$ &  $1^-$ & 15 & $\rho$\\
    \hline
     T & $\overline{\Psi^i}\gamma^0\gamma^{\mu}\Psi^j$ & $1^-$ & 15 &  $\rho$ \\
    \hline
    AV & $\overline{\Psi^i}\gamma^5\gamma^{\mu}\Psi^j$ & $1^+$ & 20 & $a_1$\\
    \hline
    AT & ~~$\overline{\Psi^i}\gamma^5\gamma^0\gamma^{\mu}\Psi^j$~~ & $1^+$ & 15 & $b_1$\\
    \hline
    \end{tabular}
\end{table}

The physical observables of interests are flavoured mesons interpolated by the fermion bilinears $\mathcal{O}_M\equiv\overline{\Psi}^i \Gamma^M \Psi^j$ with $i\neq j$. 
In \Tab{meson_ops}, we list the interpolating operators used for creating pseudoscalar (PS), scalar (S), vector (V), tensor (T), 
axial vector (AV), and axial tensor (AT) meson states 
with their assignment of spin ($J$), parity ($P$) and flavour quantum numbers as well as the corresponding names in QCD. 
Using these operators we calculate $2$-point correlation functions, 
and extract the masses and decay constants of the ground states by fitting the correlation functions to a single exponential form at later Euclidean time 
following the standard procedure. We use $Z_2\times Z_2$ single time stochastic sources, with number of hits $3$, for the measurements, see, e.g., Ref.~\cite{Bennett:2019jzz} for more details. 
As seen in \Tab{meson_ops}, vector and tensor operators interpolate the same physical state, which allows us to calculate masses for both the ground ($\rho$) and excited ($\rho'$) states 
by solving a generalised eigenvalue problem (GEVP). 
To increase the number of operator basis, we further implement smearing techniques for the link variables, APE smearing \cite{APE:1987ehd}, 
and for the fermion fields, Wuppertal smearing \cite{Gusken:1989qx}. 



\section{Results}



\begin{figure}[t]
\centering
\includegraphics[scale=0.36]{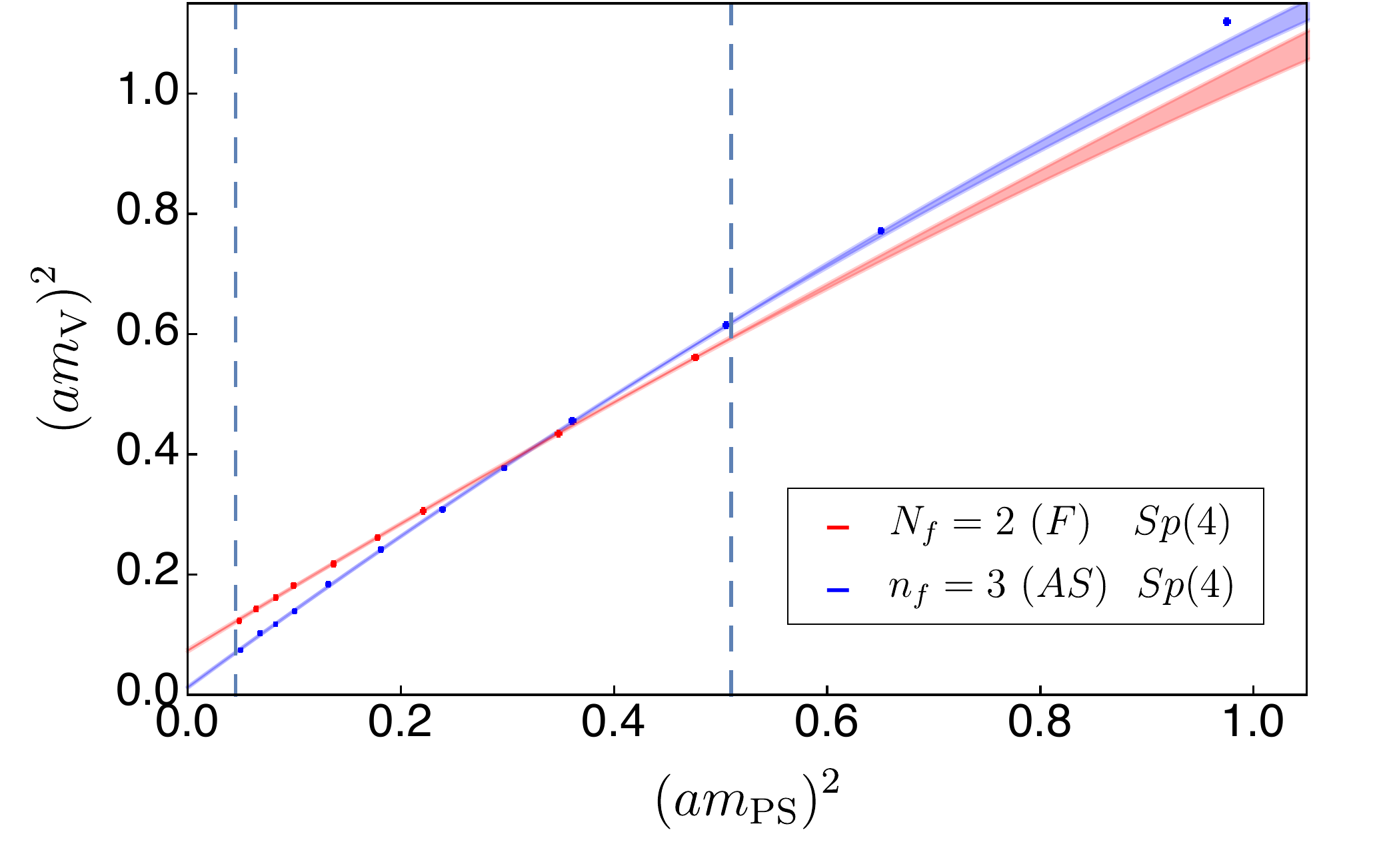}
\includegraphics[scale=0.34]{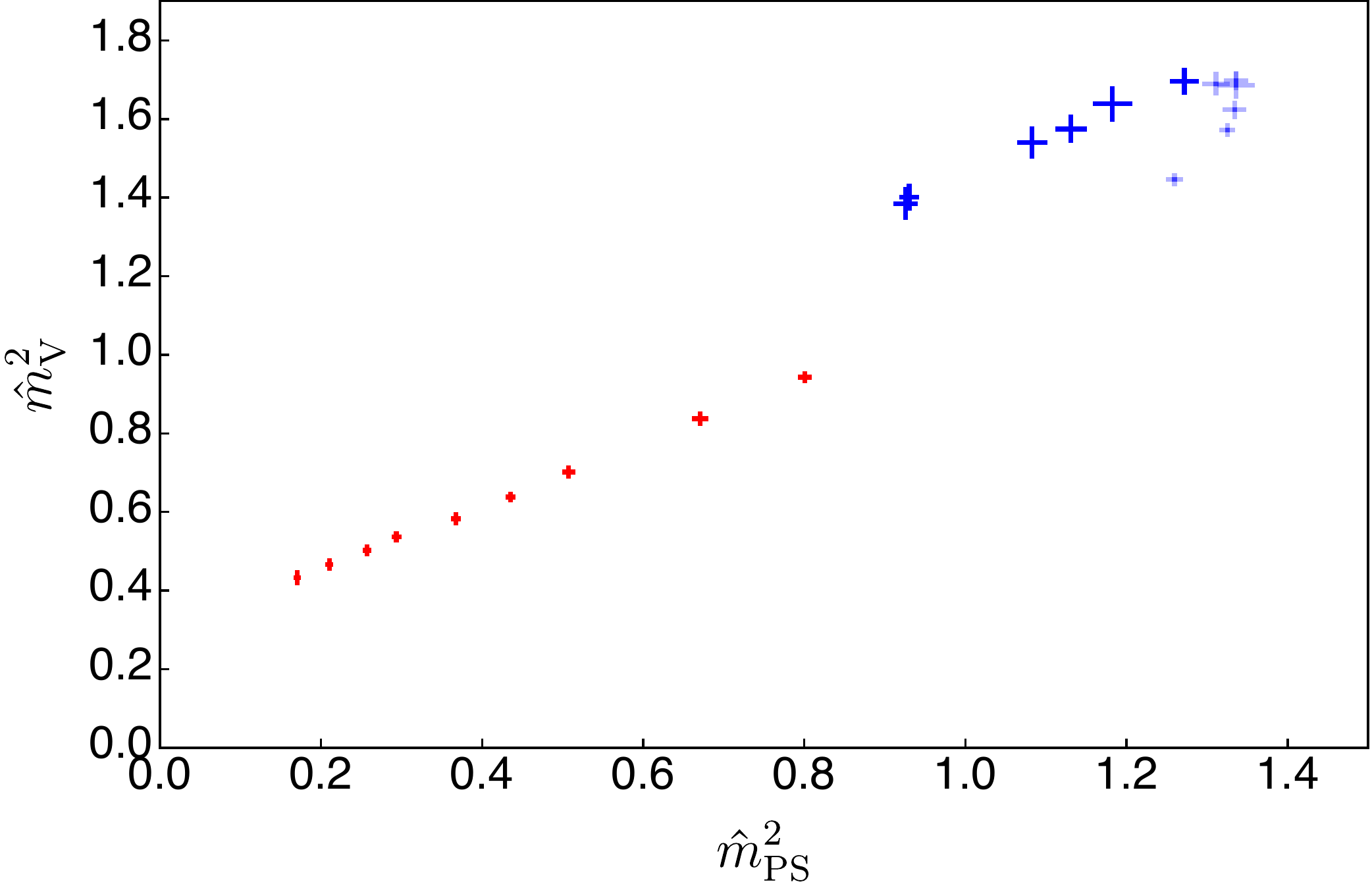}
\caption{Comparison of vector meson masses between $N_f=2$ fundamental (blue) and $n_f=3$ antisymmetric (red) $Sp(4)$ lattice gauge theories. 
The masses for the former and the latter are measured at fixed lattice couplings of $\beta=7.2$ and $6.7$, respectively. 
The left panel shows the resulting values in lattice units, while the right panel shows them in units of the gradient flow scale $w_0$. 
The coloured band in the left panel denotes the result, with $1$-$\sigma$ error, obtained from a quadratic fit to data between the two vertical dashed lines. 
}
\label{fig:m2v}
\end{figure}


Among the observables in \Tab{meson_ops}, we first focus on the pseudoscalar and vector mesons, 
the measurements of masses and decay constants of which are the most precise and highly relevant to phenomenology. 
In the left and right panels of \Fig{m2v}, we show the results of $m_{\rm V}^2$ with respect to $m_{\rm PS}^2$, 
measured at $\beta=6.7$, in lattice units and the units of $w_0$, respectively. 
For comparison purposes, we also present the results for the $Sp(4)$ theory coupled to $N_f=2$ fundamental Dirac fermions with $\beta=7.2$. 
Over the interval between the two vertical dashed lines we perform a quadratic fit to the data, and the results are denoted by the shaded lines. 
We find that the fits well describe the data over a wide region in both cases, but the extrapolated value of $(am_{\rm V})^2$ at $am_{\rm PS}=0$ are substantially different. 
If we plot the masses in units of $w_0$ the situation becomes more dramatic, as shown in the right panel of \Fig{m2v}: 
the resulting values of $\hat{m}_{\rm V}^2$ are not only residing in a small region of comparatively large mass, but also five of them, corresponding to the heavier ensembles, 
show an unexpected behaviour (light blue points). 
$w_0/a$ increases substantially as the fermion mass decreases, 
and thus results in some difficulty in the massless extrapolation. 
On the other hand, lattice artefacts affect the five aforementioned ensembles, which must be discarded in the continuum extrapolation.


\begin{figure}[t]
\centering
\includegraphics[scale=0.4]{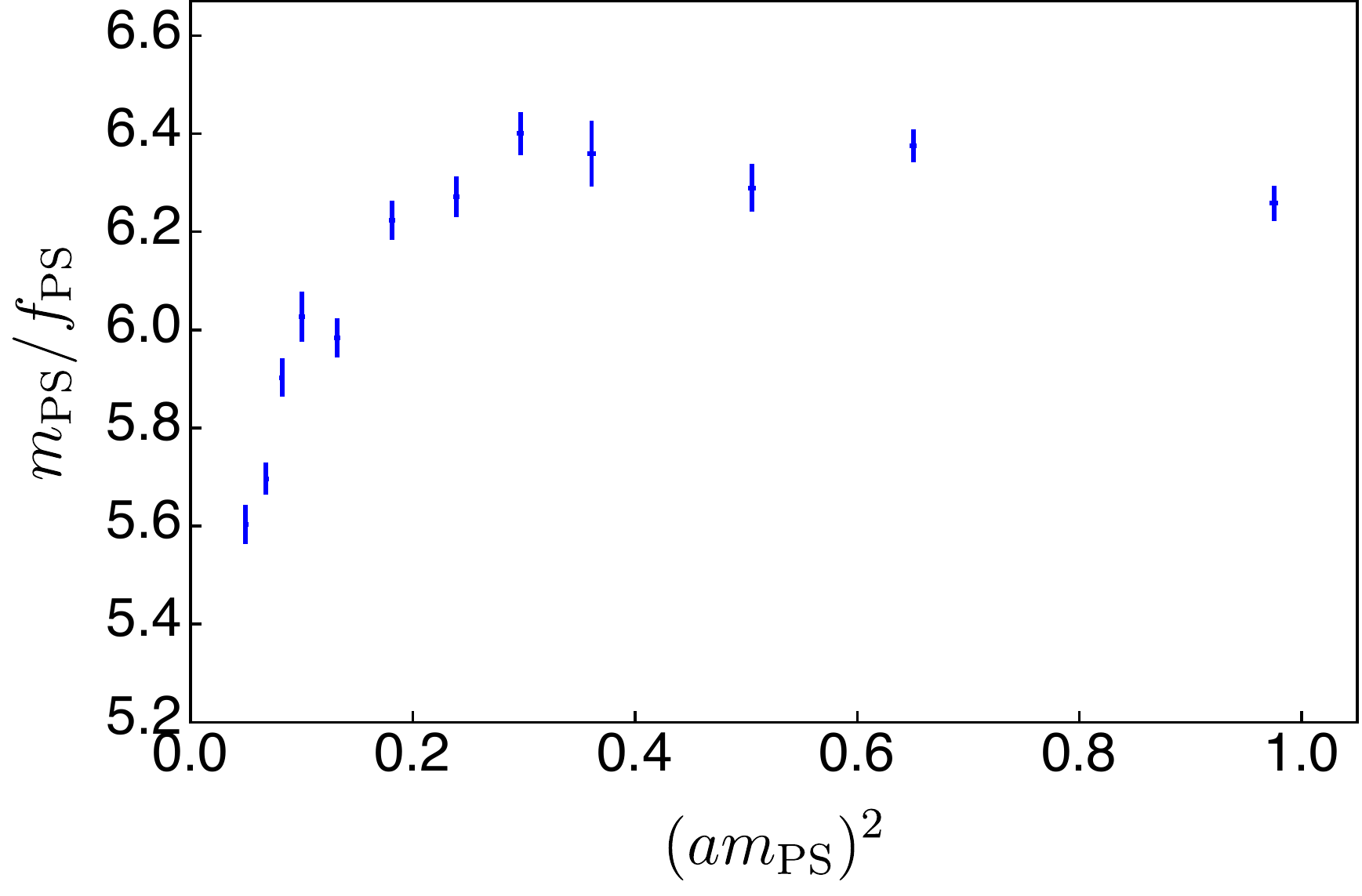}
\caption{Mass dependence of the ratio between the mass and decay constant of the pseudoscalar mesons. The lattice coupling is fixed to $\beta=6.7$. 
}
\label{fig:mpsfps}
\end{figure}

The corresponding continuum theory is expected to be well below the conformal window according to analytical calculations \cite{Lee:2020ihn}, 
which must be confirmed by nonperturbative lattice calculations. 
In conformal dynamics all dimensional quantities show hyperscaling behaviour as uniquely controlled by the fermion mass. 
We show the ratio $m_{\rm PS}/f_{\rm PS}$ measured at $\beta=6.7$ and various fermion masses in \Fig{mpsfps}. 
The relatively constant value of $m_{\rm PS}/f_{\rm PS}$ for five or six data points may be considered as a hint of near conformality. 
However, again this result is likely due to lattice artefacts, as discussed above, and we remove these measurements from the further analyses discussed for the rest of this section.  
In the small mass region, the ratio sharply decreases as $am_{\rm PS}\rightarrow 0$, which we interpret as evidence of the fact that the global symmetry is indeed broken. 

\begin{figure}[t]
\centering
\includegraphics[scale=0.357]{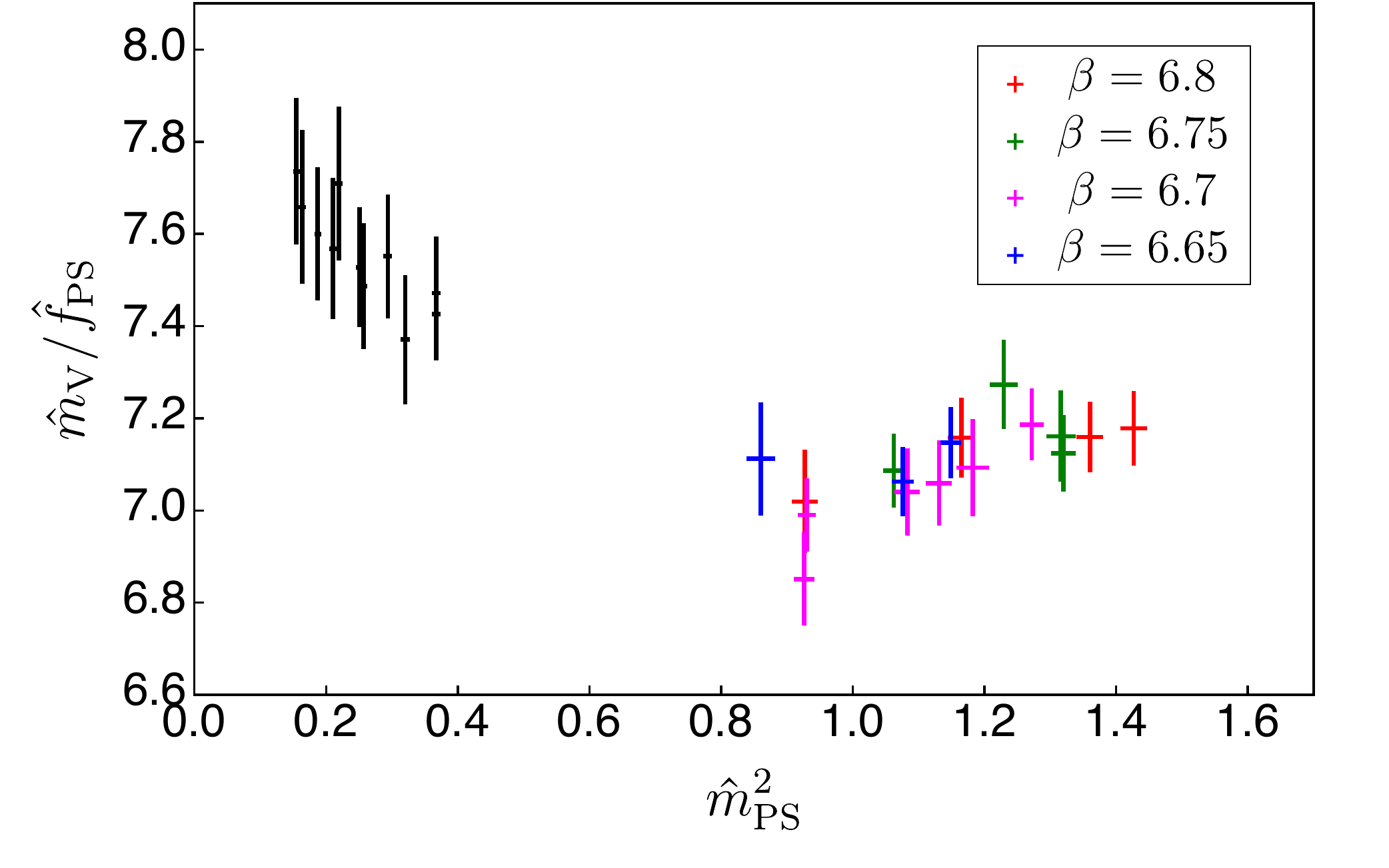}
\includegraphics[scale=0.353]{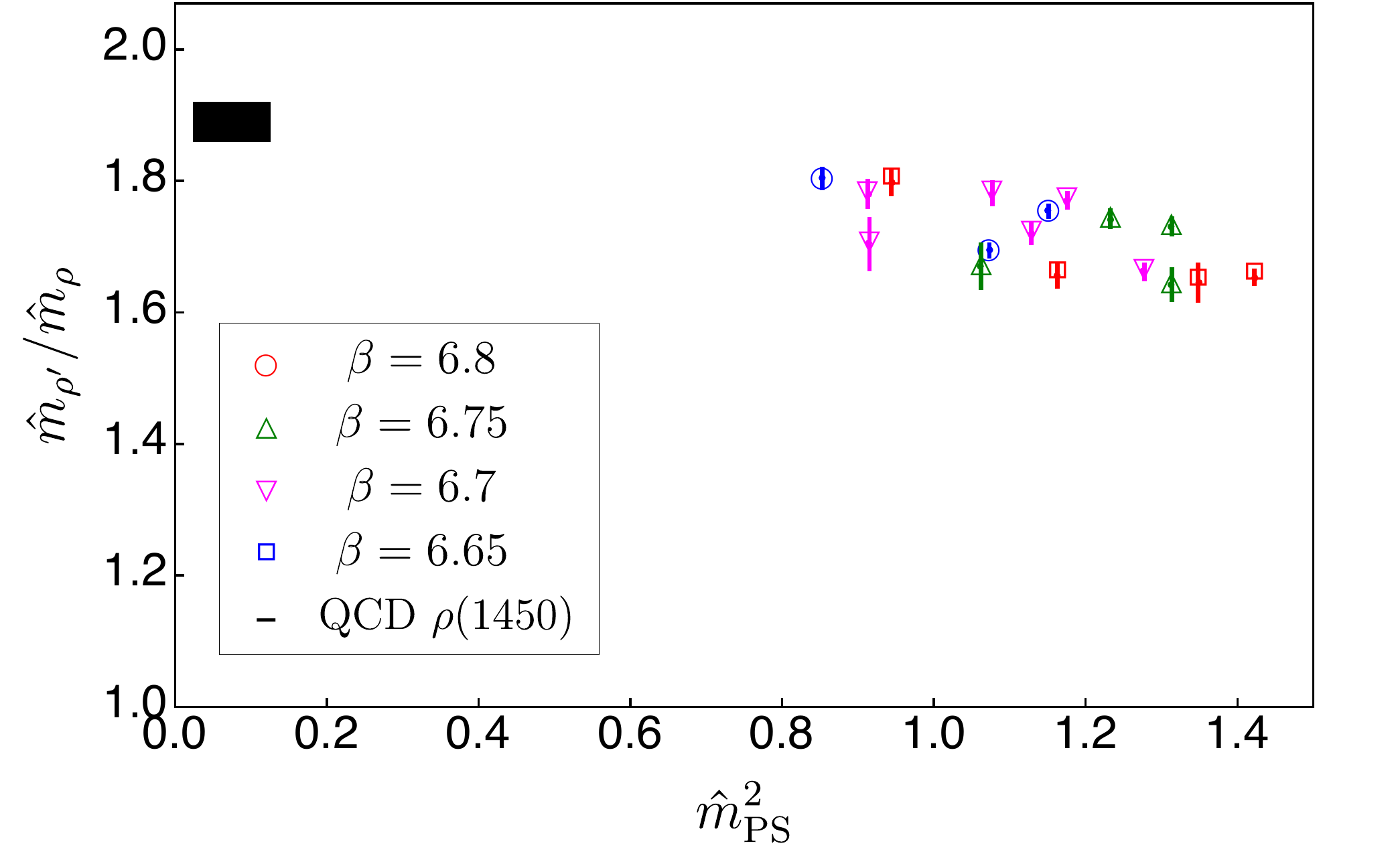}
\caption{Vector meson masses in units of the pseudoscalar decay constant (left) and mass ratios between the ground and the first excited states of vector mesons (right). 
In the left panel, for comparison purposes the continuum-extrapolated results of the $Sp(4)$ lattice gauge theory with $N_f=2$ fundamental Dirac fermions \cite{Bennett:2019jzz} are also presented in black. 
}
\label{fig:vector}
\end{figure}

We next discuss the vector meson masses. In the left panel of \Fig{vector}, we show the vector meson masses in units of the pseudoscalar decay constant measured at four different beta values, 
and compare to the continuum results for the theory with $N_f=2$ fundamental flavours. 
The resulting values of the ratio, which shows a mild dependence of both the mass and the coupling, are slightly smaller than those of the $N_f=2$ theory. 
We note that this quantity is closely related to the low-energy constant associated with the decay of vector to two pseudoscalars through the KSRF relation, 
and a comparison of various gauge theories can be found in Ref.~\cite{Bennett:2019jzz}. 
In the right panel of \Fig{vector}, we present the mass ratio between the ground and excited state vector mesons. 
Again, we are still relatively far from the massless limit and do not perform any extrapolations, and thus are unable to make a direct comparison with other theories. 
Instead, we want to emphasize that we can compute the excited states accurately using improved measurement techniques and the resulting values do not show evident anomalies. 


\begin{figure}[t]
\centering
\includegraphics[scale=0.36]{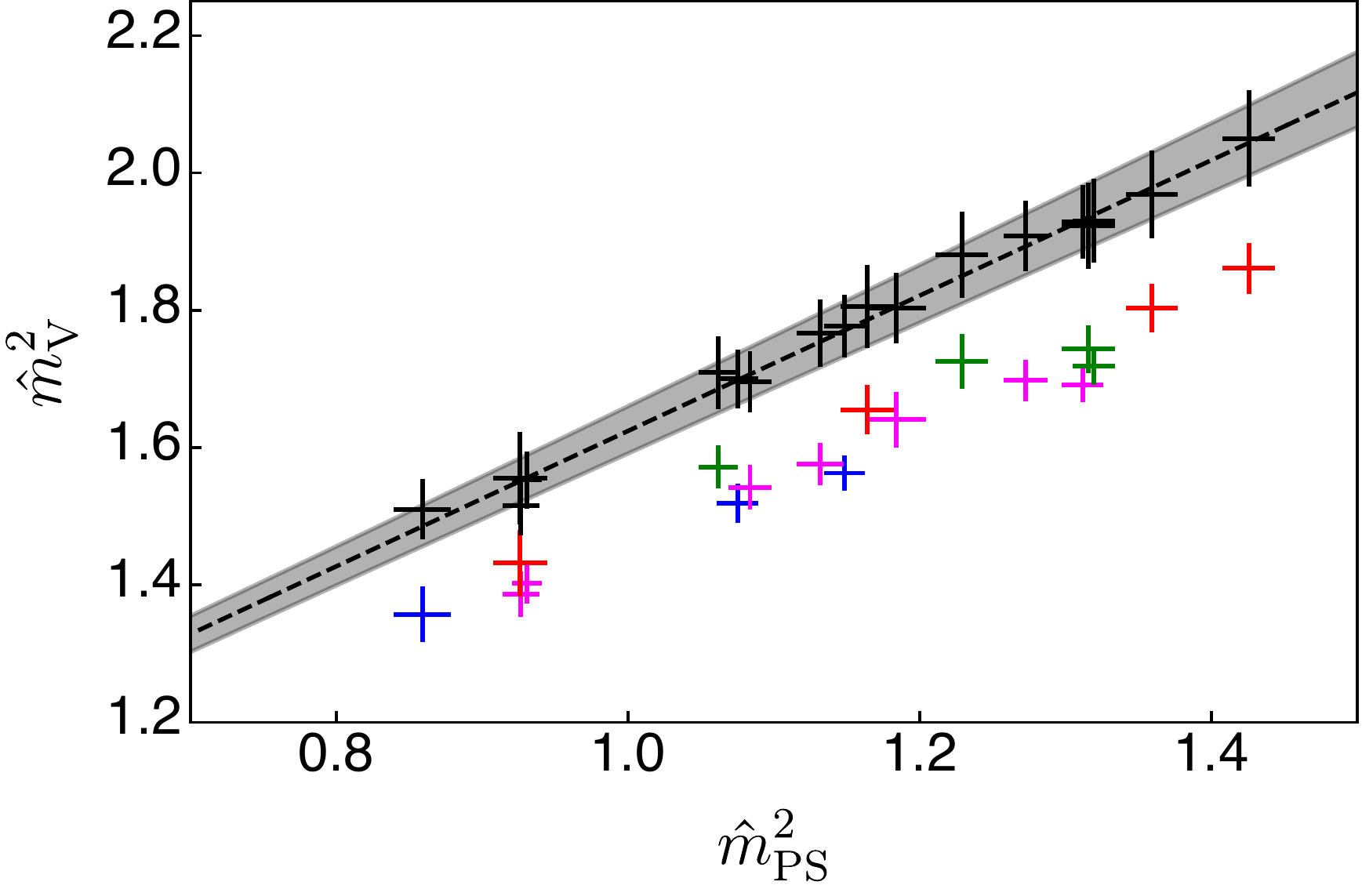}
\includegraphics[scale=0.36]{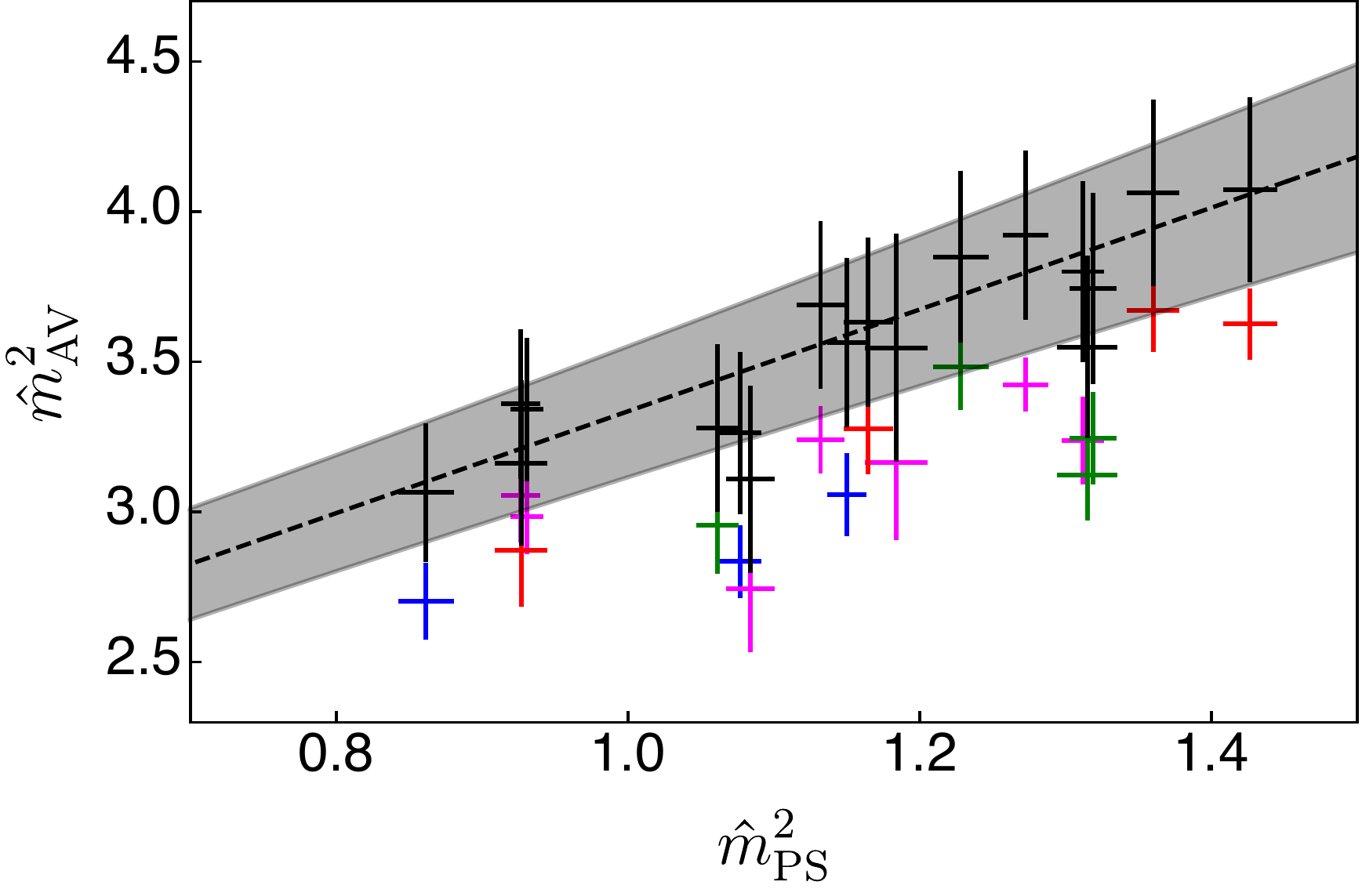}
\includegraphics[scale=0.36]{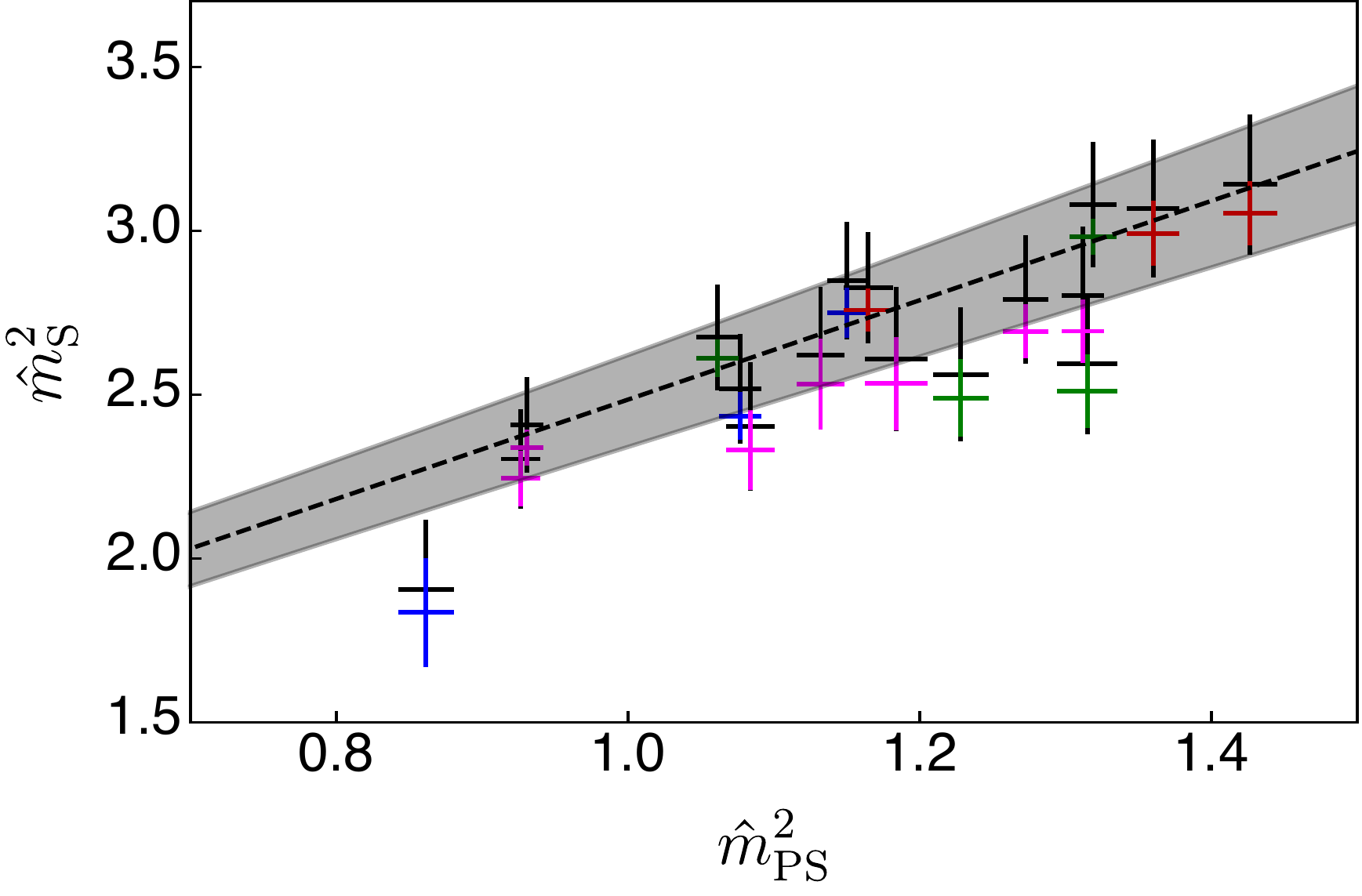}
\includegraphics[scale=0.37]{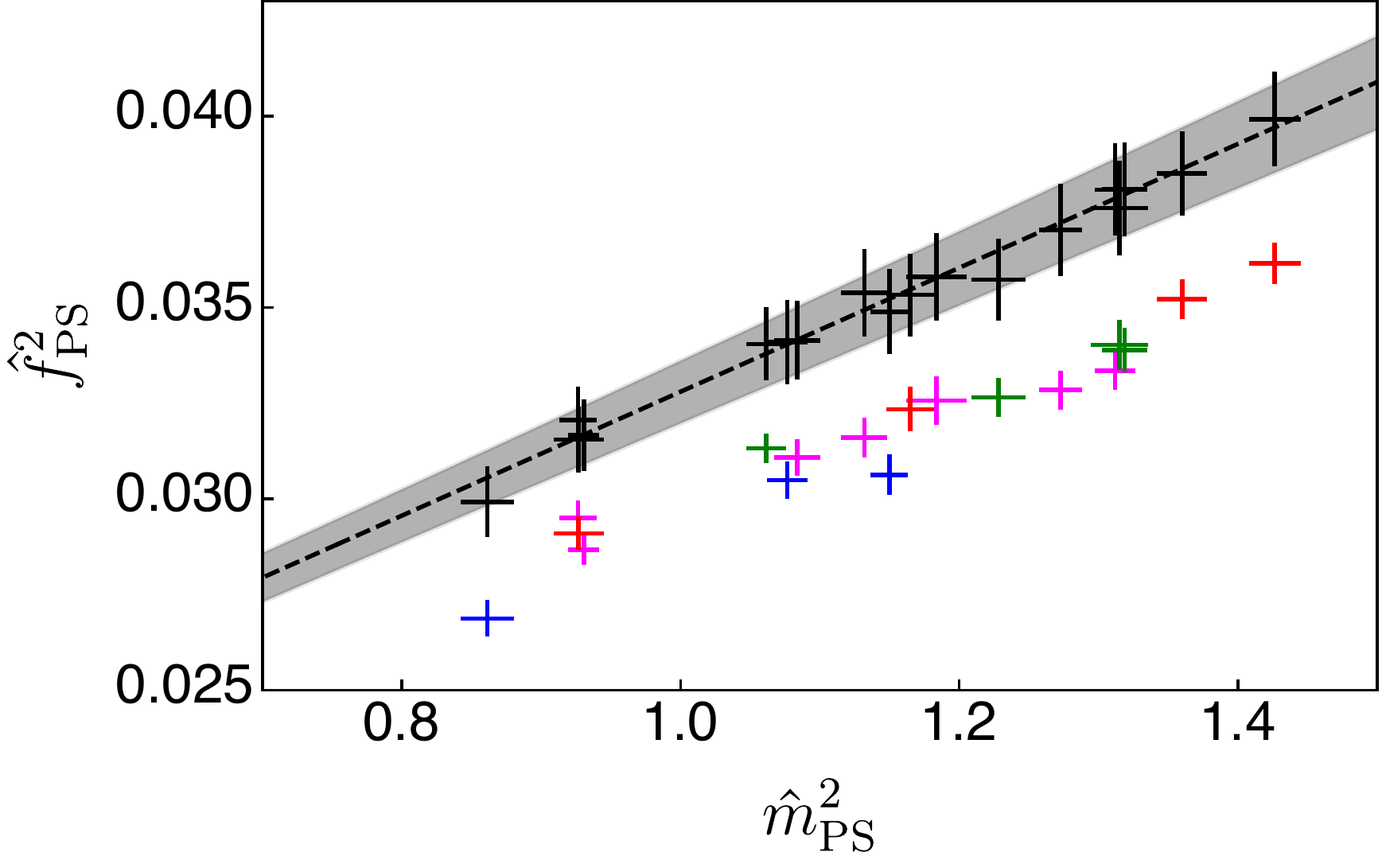}
\caption{Continuum extrapolation of meson masses and decay constants. 
Different colours denote the measurements at four different lattice couplings, $\beta=6.65$ (blue), $6.7$ (magenta), $6.75$ (green), and $6.8$ (red), 
while the black crosses denotes continuum-extrapolated results. 
}
\label{fig:meson_continuum}
\end{figure}


As discussed above, the ensembles available to us are still in a relatively large mass regime, as revealed by the scale setting with mass-dependent gradient flow method. 
We therefore do not attempt to extrapolate the results to the massless limit. Instead, we use the following simple linear ansatz to carry out the continuum extrapolation 
over the mass range showing a linear dependence of $\hat{m}_{\rm PS}^2$, 
\beqs
\hat{m}_M^{2,\,NLO} &=& \hat{m}^{2,\,\chi}_M \left(1+ L_{m,\,M} \hat{m}_{\rm PS}^2\right)+W_{m,\,M}^0 \hat{a},\\
\hat{f}_M^{2,\,NLO} &=& \hat{f}^{2,\,\chi}_M \left(1+ L_{f,\,M} \hat{m}_{\rm PS}^2\right)+W_{f,\,M}^0 \hat{a},
\eeqs
where $\hat{m}^{\chi}_M$, $\hat{f}^{\chi}$, $L_M$ and $W^0$ are the low-energy constants determined from the fits. 
In \Fig{meson_continuum}, we show the results of the continuum fit, which are denoted by grey bands, along with the measurements at finite lattice spacing 
and the continuum-extrapolated ones after subtracting the finite-spacing corrections. 
We only show preliminary results for the mass of vector, axial-vector and scalar mesons and the decay constant of pseudoscalar mesons. 
Full results will be presented in our future publication by considering all the measurements along with more dedicated analysis.


\section{Summary}

We studied the mass spectrum of spin-$0$ and $1$ mesons, including the first excited state of vector meson, for the $Sp(4)$ gauge theory with three antisymmetric Dirac fermions on the lattice 
using the standard plaquette action and Wilson-Dirac fermions. 
For pseudoscalar, vector and axial vector mesons we also calculated the decay constants. 
Our first observation is that the ratio $m_{\rm PS}/f_{\rm PS}$ is not a constant, but shows a sharp drop towards the massless limit, implying that the theory is in the broken phase. 
We found a significant mass dependence of the gradient flow scale, 
which in turn presents us with challenges in approaching the massless limit. 
We carried out the continuum extrapolation using a linear ansatzs in the pseudoscalar mass squared and the lattice spacing. 
Our results provide useful nonperturbative input for phenomenological studies 
of composite Higgs models. 

\acknowledgments

The work of J.~W.~L is supported by the National Research Foundation of Korea (NRF) grant funded by the Korea government (MSIT)
(NRF-2018R1C1B3001379).
The work of E. B. has been funded in part by the UKRI Science and
Technology Facilities Council (STFC) Research Software Engineering Fellowship EP/V052489/1. 
The work of D.~K.~H. was supported by
Basic Science Research Program through the National
Research Foundation of Korea (NRF) funded by the
Ministry of Education (NRF-2017R1D1A1B06033701).
The work of H.~H. and C.~J.~D.~L. is supported
by the Taiwanese MoST Grant No. 109-2112-M-009 -006 -MY3. The work of B.~L. and M.~P. has been supported in part
by the STFC Consolidated Grants No. ST/P00055X/1 and No. ST/T000813/1. B.~L. and M.~P. received funding from
the European Research Council (ERC) under the European
Union’s Horizon 2020 research and innovation program
under Grant Agreement No. 813942. The work of B.~L. is
further supported in part by the Royal Society Wolfson
Research Merit Award No. WM170010 and by the
Leverhulme Trust Research Fellowship No. RF-2020-4619. 
The work of D.~V. is supported in part the Simons Foundation under the program “Targeted Grants to Institutes” awarded to the Hamilton Mathematics Institute.
Numerical simulations have been performed on the Swansea SUNBIRD
cluster (part of the Supercomputing Wales project) and AccelerateAI A100 GPU system,
on the local HPC clusters in Pusan National
University (PNU) and in National Yang Ming Chiao Tung University
(NYCU), and on the DiRAC Data Intensive service at Leicester. 
The Swansea SUNBIRD system and AccelerateAI are part funded
by the European Regional Development Fund (ERDF) via
Welsh Government. The DiRAC Data Intensive service at Leicester is operated 
by the University of Leicester IT Services, which forms part of the STFC DiRAC HPC Facility (www.dirac.ac.uk). The DiRAC Data Intensive service equipment at Leicester was funded by BEIS capital funding via STFC capital
grants ST/K000373/1 and ST/R002363/1 and STFC DiRAC Operations grant ST/R001014/1. 
DiRAC is part of the National e-Infrastructure.

For the purpose of open access, the authors have applied a Creative Commons Attribution (CC BY) licence to any Author Accepted Manuscript version arising.

\end{document}